\documentclass[aps,prl,reprint,floatfix,superscriptaddress,amsmath,showpacs]{revtex4-1}

\usepackage{graphicx}
\usepackage{amsmath}
\usepackage{textcomp} 

\begin{document}

\title{Sub-microsecond fast temporal evolution of the spin Seebeck effect}

\author{M. Agrawal}
\email{magrawal@physik.uni-kl.de}
\affiliation{Fachbereich Physik and Landesforschungszentrum OPTIMAS, Technische Universit\"at Kaiserslautern, 67663 Kaiserslautern, Germany}
\affiliation{Graduate School Materials Science in Mainz, Gottlieb-Daimer-Strasse 47, 67663 Kaiserslautern, Germany }

\author{V. I. Vasyuchka}
\affiliation{Fachbereich Physik and Landesforschungszentrum OPTIMAS, Technische Universit\"at Kaiserslautern, 67663
Kaiserslautern, Germany}

\author{A. A. Serga}
\affiliation{Fachbereich Physik and Landesforschungszentrum OPTIMAS, Technische Universit\"at Kaiserslautern, 67663
Kaiserslautern, Germany}

\author{A. Kirihara}
\affiliation{Fachbereich Physik and Landesforschungszentrum OPTIMAS, Technische Universit\"at Kaiserslautern, 67663 Kaiserslautern, Germany}
\affiliation{Smart Energy Research Laboratories, NEC Corporation, Tsukuba 305-8501, Japan}

\author{P. Pirro}
\affiliation{Fachbereich Physik and Landesforschungszentrum OPTIMAS, Technische Universit\"at Kaiserslautern, 67663 Kaiserslautern, Germany}

\author{T. Langner}
\affiliation{Fachbereich Physik and Landesforschungszentrum OPTIMAS, Technische Universit\"at Kaiserslautern, 67663
Kaiserslautern, Germany}

\author{M. B. Jungfleisch}
\affiliation{Fachbereich Physik and Landesforschungszentrum OPTIMAS, Technische Universit\"at Kaiserslautern, 67663
Kaiserslautern, Germany}

\author{ A. V. Chumak}
\affiliation{Fachbereich Physik and Landesforschungszentrum OPTIMAS, Technische Universit\"at Kaiserslautern, 67663
Kaiserslautern, Germany}

\author{E. Th. Papaioannou}
\affiliation{Fachbereich Physik and Landesforschungszentrum OPTIMAS, Technische Universit\"at Kaiserslautern, 67663
Kaiserslautern, Germany}

\author{B. Hillebrands}
\affiliation{Fachbereich Physik and Landesforschungszentrum OPTIMAS, Technische Universit\"at Kaiserslautern, 67663 Kaiserslautern, Germany}

\date{\today}

\begin{abstract}

We present temporal evolution of the spin Seebeck effect in a YIG$\vert$Pt bilayer system. Our findings reveal that this effect is a sub-microseconds fast phenomenon governed by the temperature gradient and the thermal magnons diffusion  in the magnetic materials. A comparison of experimental results with the thermal-driven magnon-diffusion model shows that the temporal behavior of this effect depends on the time development of the temperature gradient in the vicinity of the YIG$\vert$Pt interface. The effective thermal-magnon diffusion length for YIG$\vert$Pt systems is estimated to be around 700\,nm.

\end{abstract}

\maketitle

The spin Seebeck effect (SSE) \cite{Uchida2008,Uchida2011,Jaworski2010,Jaworski2011,Uchida2010,Uchida2010a,Adachi2010,Uchida2010c} is one of  the most fascinating phenomena in the contemporary era of spin-caloritronics\cite{Bauer2012}. Analogous to the classical Seebeck effect, the SSE is a phenomenon where a spin current is generated in spin-polarized materials like metals \cite{Uchida2008}, semiconductors \cite{Jaworski2010,Jaworski2011}, and insulators \cite{Uchida2010,Uchida2010a,Adachi2010,Uchida2010c} on the application of a thermal gradient. Generally, the generated spin current is measured by the  inverse spin Hall effect (ISHE) \cite{Saitoh2006}  in a normal metal like Pt, placed in contact with the spin-polarized material. Currently, this phenomenon has attracted much attention due to its potential applications, for example, recent progresses show that based on this effect  thin-film structures can be fabricated to generate electricity from waste-heat sources \cite{Kirihara2012}. Further advancements in industrial applications like temperature sensors, temperature gradient sensors, and thermal spin-current generators require an in-depth understanding of this effect.

Although there have been numerous  experimental and theoretical studies about this effect, the underlying physics is yet not well understood. The most accepted theory predicts that the SSE is driven by the difference in the local temperatures of  magnon-, phonon-, and electron baths \cite{Xiao2010, Ohe2011} of the system. However, no clear evidences of such differences have been observed experimentally \cite{Agrawal2013}. So, the origin of this effect is still under discussion. Some experimental studies show that the interface proximity effect in the YIG$\vert$Pt system could exhibit similar behavior as observed for the SSE \cite{Huang2012}. However, very recent measurements claim no such proximity effects \cite{Geprags2012}. Moreover, the question whether the SSE is an interface or bulk effect, is still open \cite{Kehlberger2013,Schreier2013}.  

To shed light on this controversial physics, we developed an entirely new experimental approach where we studied the temporal evolution of the SSE in YIG$\vert$Pt bilayer structures. The observations were realized in the longitudinal configuration of the SSE \cite{Uchida2010a,Uchida2010c}. In the longitudinal spin Seebeck effect (LSSE), a thermal gradient is created perpendicular to the film plane, and the spin current generated by thermal excitations of magnetization (thermal magnons) is measured along the thermal gradient. From our measurements, we find that the SSE signal evolves at sub-microsecond time-scales, and a certain thickness of the YIG film effectively contributes to the SSE.

\begin{figure}[t]
\includegraphics[width=8 cm]{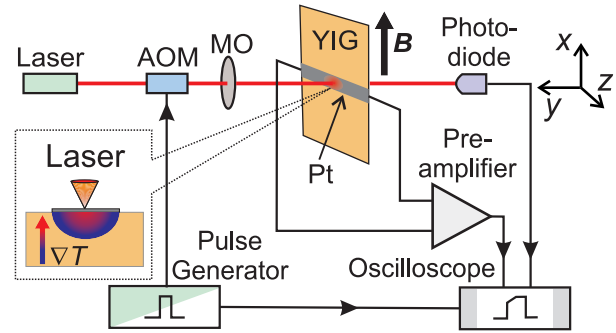}
\caption{\label{Fig1} (Color online) Sketch of the experimental setup. A continuous laser beam (wavelength 655\,nm), modulated by an acousto-optical modulator (AOM), was focused down on a 10\,nm thick Pt strip, deposited on a 6.7\,$\mu$m thick YIG film, by a microscope objective (MO). The laser intensity profile was monitored by an ultrafast photo-diode. An in-plane magnetic field $B=20\,mT$ was applied to the YIG film. The heated Pt strip created a thermal gradient perpendicular to the YIG$\vert$Pt interface (see the inset). The generated  voltage across the Pt strip due to ISHE was amplified and measured by an oscilloscope.}
\end{figure}

\begin{figure}[t]
\includegraphics[width=8.5 cm]{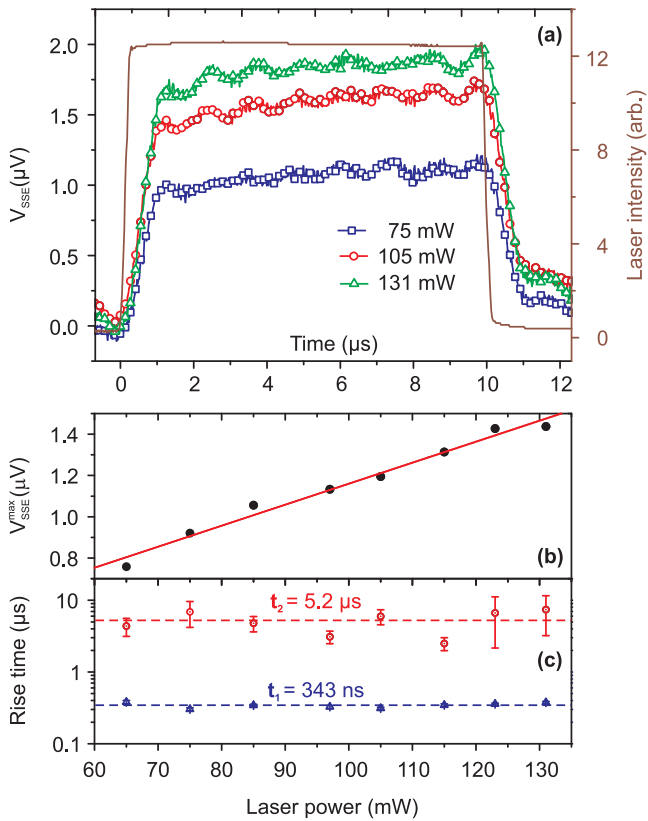}
\caption{\label{Fig2} (Color online) (a) The  time profiles of the laser intensity (solid line), and the SSE voltage ($V_\mathrm{SSE}$) at various laser powers of 75\,mW (open circle), 105\,mW (open square), and 131\,mW (open triangle). (b) Measured $V_\mathrm{SSE}^\mathrm{max}$, with linear fitting, and  (c) the rise times $t_\mathrm{1}$ and $t_\mathrm{2}$ as a function of laser power. The rise times are practically unchanged with laser power.}
\end{figure}

The LSSE measurements were performed  on a bilayer of magnetic insulator, Yttrium Iron Garnet (YIG), and normal metal, Pt. A 6.7\,$\mu$m thick YIG sample of dimensions 14\,mm$\times$3\,mm was grown by liquid phase epitaxy on a 500\,$\mu$m thick Gallium Gadolinium  Garnet (GGG)  substrate. To achieve a good YIG$\vert$Pt interface quality, a detailed cleaning process as discussed in Ref. \cite{Jung2013} was followed before the deposition of Pt. A strip (3\,mm$\times$100 $\mu$m) of 10\,nm thick Pt  was deposited on the cleaned YIG surface by MBE at a pressure of $5 \times10^{-11}$\,mbar with a growth rate of 0.05\,nm/s. In Fig.~\ref{Fig1}, a schematic diagram of the experimental setup is shown. A laser heating technique \cite{Walter2011,Boehenke2013} was implemented to heat the Pt strip from the top surface to create a vertical thermal gradient  along the \textit{y} direction perpendicular to the bilayer interface (See Fig.\,\ref{Fig1}). For this purpose, a continuous laser beam (wavelength 655\,nm) was modulated by an acousto-optical modulator (AOM), and focused down at the middle of the Pt strip using a microscopic objective (Leitz PL 16\,x/0.30). To study the temporal profile of the laser beam in parallel, the transmitted laser beam through the YIG sample was monitored by an ultrafast photo-diode. A (10\%-90\%)  rise time of  200\,ns was observed for the laser pulses (solid line in Fig.\ref{Fig2}(a)).   The sample structure was mounted on a copper block to provide a thermal heat sink.

The time-resolved measurements of LSSE were carried out using a 10\,$\mu$s long laser pulse with a repetition rate of 10\,kHz.  An in-plane magnetic field $B=20$\,mT was applied to saturate the YIG film magnetization along the \textit{x} direction. As a result of the LSSE, a spin current flowed along the \textit{y} direction. By the inverse spin Hall effect, this spin current converts into an electric field along the \textit{z} direction in Pt. The electric field was detected as a potential difference $V_\mathrm{SSE}$ between the two short edges of the Pt strip (shown in Fig.\,\ref{Fig1}). The SSE voltage $V_\mathrm{SSE}$ was amplified by a high input impedance preamplifier and monitored on an oscilloscope.The measurements were performed for both  $\pm$\,\textit{x} directions of the magnetic field. The SSE voltage changes its polarity by reversing the direction of magnetic field \cite{Uchida2010a}; an absolute average value of $V_\mathrm{SSE}$ was evaluated to eliminate the thermal emf offset.

In Fig.~\ref{Fig2}(a), the temporal profile of the laser light intensity and $V_\mathrm{SSE}$ for different laser heating powers are plotted. The SSE signal rises sharply for the first 1$\mu$s and then gradually attains a saturation level $V_\mathrm{SSE}^\mathrm{max}$. With increasing laser power,  $V_\mathrm{SSE}^\mathrm{max}$ increases linearly (Fig.~\ref{Fig2}(b)). This linear behavior indicates that the laser heating is in the linear regime, and no nonlinear phenomena are involved in this process. A comparison of the rising edges of the laser intensity and the SSE signal provides a clear signature that the SSE has no direct correlation with the laser intensity profile.

To understand the ongoing mechanism, we first analyzed  the rise times of $V_\mathrm{SSE}$  for different laser powers. We fitted $V_\mathrm{SSE}$ with a saturating  double exponential function ($1-A\exp(-t/t_1)-B\exp(-t/t_2)$) which essentially correlates with the heat dynamics of a system owing to heat losses. In Fig.~\ref{Fig2}(c), the rise times, $t_1$ and $t_2$, show that, within the limits of experimental error, they are independent of the laser power. Average values of $t_\mathrm{1}=343\,\mathrm{ns}$ and $t_\mathrm{2}=5.2\,\mu\mathrm{s}$ were obtained by using data shown in Fig.~\ref{Fig2}(c). These rise times are much different from the rise time  of the laser intensity (10\%-90\% rise time $\approx\,$200\,ns).

\begin{figure}[b]
\includegraphics[width=8.5 cm]{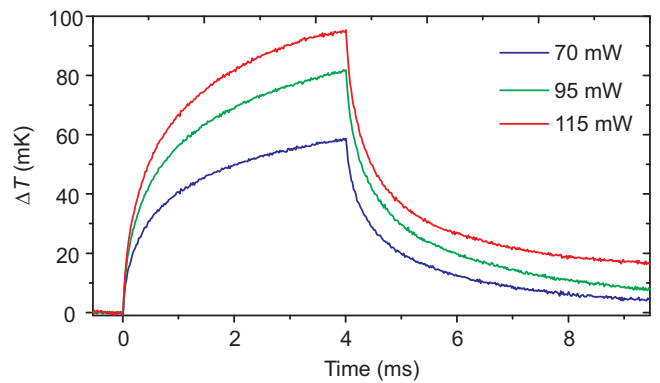}
\caption{\label{Fig3} (Color online) The time profile of the variation of temperature in Pt on heating with a 4\,ms long laser pulse with various powers.}
\end{figure}

Our first hunch to interpret these rise times was to study the temperature evolution in the YIG$\vert$Pt system. Fortunately, the Pt strip grown over the YIG film can be utilized as a perfect resistance-temperature-detector to measure the temperature at the surface of the YIG film. We performed the resistance measurements of the Pt strip to calculate the variation of the temperature in the YIG$\vert$Pt system by the laser heating. To do so, a constant current $I_\mathrm{c}=0.5$\,mA was passed through the Pt strip (room temperature resistance $R_\mathrm{Pt}\approx 508\,\Omega$), and the potential drop ($\Delta V_\mathrm{Pt}=\Delta R_\mathrm{Pt} I_\mathrm{c}$) due to the heating of Pt  was measured with the same experiment setup used for the $V_\mathrm{SSE}$ measurements. A much longer laser pulse of 4\,ms  was used to heat the Pt strip. Note that the thermal emf (few microvolts) has negligible influence on these measurements as the potential drop ($\Delta V_\mathrm{Pt}$) is very large ($\approx\,0.25\,\mathrm{V}$). With the help of  auxiliary measurements of static resistance versus temperature performed on the same Pt strip,  $\Delta V_\mathrm{Pt}$ can be expressed in terms of temperature ($T$).  In Fig.~\ref{Fig3}, the variation in the  temperature of the Pt strip  ($\Delta T$) is plotted for different laser powers. A rise time of 2\,ms, obtained by fitting the data with a single saturating exponential function, is three orders of magnitude longer than the rise time of the SSE signal. The saturating exponential behavior of the temperature illustrates that the heat losses in the system dominantly control the heat dynamics of Pt. Further, measurements evident  that, likewise the SSE signal,  the rise time of the temperature is also independent of the laser power.

Clearly, from the measurement of the Pt resistance, it can be concluded that the temperature of the system has no direct correlation with the fast time-scale of the SSE. To dig out the cause attributing to the fast rising of the SSE, we propose a model  where we consider the thermally-induced motion of magnons  in  a system of normal metal$\vert$magnetic material (e.g., Pt$\vert$YIG) subject to a thermal gradient. In such a system, the spin current flowing in/out of the normal metal depends upon the temperature difference of the magnon- and the phonon baths at the interface \cite{Xiao2010,Ohe2011} and the magnon accumulation close to the interface in the magnetic material \cite{Kehlberger2013}. On the application of a temperature gradient,  thermal magnons having higher population at hotter regions---in equilibrium their population is proportional to the phonon temperature---propagate towards colder regions with less magnon population. The propagation of magnons creates a magnon density gradient in the  system along with the phonon thermal gradient. This implies that the spatial distribution of the magnon density depends on the magnon population (phonon temperature) and their propagation lengths. Therefore, the spin Seebeck voltage can be considered as a combination of an interface effect and a bulk contribution from the magnon motions and, eventually,  can be expressed as
\begin{equation}
\label{eq1}
 V_\mathrm{SSE}\propto \alpha (T_N-T_M) + \beta \int \limits_{y} \nabla T_y \exp(-y/L)\,dy \, ,
\end{equation}  
where $T_N$ is the phonon temperature (= electron temperature) in the normal metal, $T_M$ the magnon temperature at the interface, $\nabla T_y$  the phonon thermal gradient perpendicular to the interface, and \textit{L} the effective magnon diffusion length. The parameter $\alpha$ defines the coupling between the electron bath in the normal metal and the magnon bath in the  magnetic material. The coupling parameter $\beta$ specifies the magnon-magnon coupling within the magnetic material. The second term of Eq.\,(\ref{eq1}) is an integration along the phonon thermal gradient over the thickness of the magnetic material.

\begin{figure}[h]
\includegraphics[width=8.5 cm]{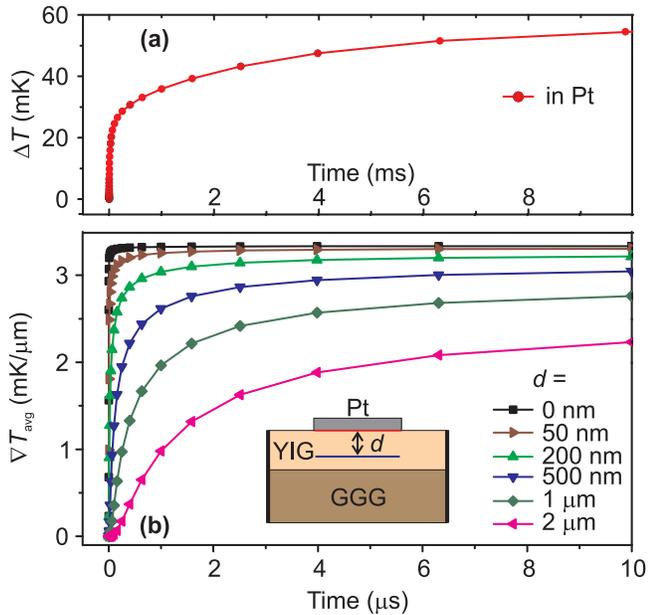}
\caption{\label{Fig4} (Color online) Numerically calculated time profile of (a) the temperature in Pt and (b) thermal gradients, $\nabla T_\mathrm{avg}$, in YIG  at \textit{d}\,=\,0\,nm (interface), 50\,nm, 200\,nm, 500\,nm, 1\,$\mu$m, and 2$\,\mu$m distances away form the YIG$\vert$Pt interface.The inset shows the model geometry of the COMSOL simulations. The thick vertical lines represent the constant temperature  boundaries (293.15\,K).}
\end{figure}

In order to determine the phonon thermal gradient $\nabla T_y$ , we numerically solved the 2D phonon heat conduction equation  for the  YIG$\vert$Pt bilayer using the COMSOL Multiphysics simulation package \cite{comsol}. In the simulation model, a 10\,nm thick and 10\,$\mu$m wide Pt rectangular block was placed on a 6.7\,$\mu$m thick and 300\,$\mu$m wide YIG film. The entire structure was mounted on a GGG substrate (50\,$\mu$m$\times$300\,$\mu$m). The simulation parameters are indicated in Table \ref{table1}. The YIG$\vert$Pt interfacial thermal resistance \cite{Schreier2013} was implemented in the simulations. However, this consideration made no remarkable difference in the outcome of the simulations. As boundary conditions, the temperatures along the short edges of the YIG and GGG layers  (see the inset in Fig.~\ref{Fig4}(b)) were kept fixed to 293.15\,K. These boundaries resembled the heat sink in the experimental set-up. A 2\,$\mu$m wide area at the middle of the Pt block was considered as a heat source which replicated the laser heating in the experimental set-up.

\begin{table}
\caption{\label{table1} Material parameters used for the numerical solution of the phonon heat transport equations in  the YIG$\vert$Pt system.}
\begin{ruledtabular}
\begin{tabular}{c c c c }
material & density & thermal conductivity & heat capacity \\
 & (kg/$\mathrm{m^{3}}$) & (W/m K) &  (J/kg K) \\
\hline
Pt & 21450 \footnotemark [1] & 20 \footnotemark [2] & 130 \footnotemark [1]\\
YIG & 5170 \footnotemark [4] & 6.0 \footnotemark [3] & 570 \footnotemark [5]\\
GGG & 7080 \footnotemark [3] & 7.94\footnotemark [3] & 405\footnotemark [3] \\

\end{tabular}
\end{ruledtabular}

\footnotetext [1] {Ref. \cite{Lide2008}}  \footnotetext [2] {Ref. \cite{Zhang2006}}  \footnotetext [3] {Ref. \cite{Hofmeister2006}}
\footnotetext [4] {Ref. \cite{Clark1961}} \footnotetext [5] {Ref. \cite{Guillot1981}} 

\end{table}

In Fig.~\ref{Fig4}(a), the simulated temporal evolution of the average temperature in Pt is shown. The temperature dynamics in Pt ($\approx$\,3\,ms) was obtained as slow as it was observed in the Pt-resistance-measurement experiment. From simulations, we find that a gradual increase in the average temperature is due to the large- heat capacity and volume of the system.  On the other hand, the thermal gradient close to interface shows fast dynamics.  We evaluated the average thermal gradient $\nabla T_\mathrm{avg}$ along lines parallel to interface  for various distances \textit{d} from the interface in the YIG film (see the inset in Fig.~\ref{Fig4}(b)). These parallel lines essentially represent the parallel planes (\textit{xz}) in the experimental geometry. In Fig~\ref{Fig4}(b), the average thermal gradient for different distances \textit{d} from the interface are shown. Contrary to the temperature in Pt, the average thermal gradient rises very rapidly and saturates within microseconds. As \textit{d} , i.e., the depth of the reference line (a plane in 3D model) from the interface, increases, the rise time of the temperature gradient raises due to the slow-down of the heat flow caused by a finite thermal conductivity and the increasing thermal capacity ($\propto$ volume). Note that even after the first 10$\mu$s, the heat was not fully distributed up to the ends of the Pt block, therefore the dimensions of the model  were not affecting the temporal profile of thermal gradients. Furthermore, the simulations show that the lateral heat flow in the YIG film and the heat transport within the Pt strip have minor influences on the average thermal gradients.

\begin{figure}[t]
\includegraphics[width=8.5 cm]{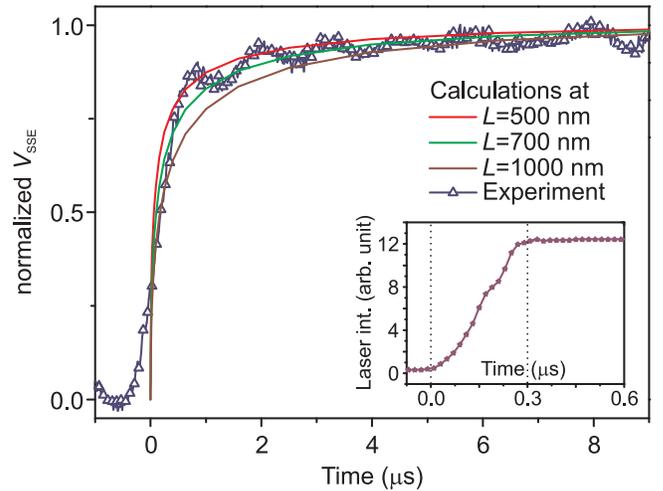}
\caption{\label{Fig5} (Color online) Comparison of normalized spin Seebeck voltage $V_\mathrm{SSE}$ measured experimentally with the numerical calculations for different effective magnon   diffusion lengths \textit{L}\,=\,500\,nm, 1\,$\mu$m, 2\,$\mu$m. The inset shows the switching time ($\approx\,0.25\,\mu\mathrm{m}$) of the laser intensity.}
\end{figure}

The fast rise of the thermal gradient ($\approx\,$50\,ns) at the interface of the YIG$\vert$Pt bilayer (\textit{d}\,=\,0\,nm), shown in Fig.~\ref{Fig4}(b), leads to a conclusion that the time-scale  of the SSE cannot be explained  by examining only the time evolution of the thermal gradient at the interface. The time scale of the SSE must be influenced by a rather slower process. On the basis of this argument, the first term of Eq.\,(\ref{eq1}), which is proportional to the phonon thermal gradient at the interface \cite{Xiao2010,Schreier2013}, can be considered static over the time-scale of our interest ($>$\,50\,ns). Using the phonon thermal gradient data, obtained from the COMSOL simulations, we calculated the integral term of Eq.\,\ref{eq1} for different magnon propagation lengths. The integral was computed from the interface up to the thickness of the YIG film.
In Fig.~\ref{Fig5}, the normalized value of the experimentally and  numerically calculated  $V_\mathrm{SSE}$ for \textit{L}\,=\,500\,nm,~700\,nm, and 1000\,nm are plotted as a function of  time. Clearly, our model replicates the experimentally observed time scales of the SSE. On comparing the calculated $V_\mathrm{SSE}$ with the experimental results, we estimate the effective magnon diffusion length to be equal to $\approx$ 700\,nm. Note that the very first slow increase in the normalized $V_\mathrm{SSE}$ (for time\,$\leq\,0\,\mu$s) is originated from the switching time of the laser ($\approx~0.25\,\mu\mathrm{m}$), shown in the inset to Fig~\ref{Fig5}. The effective magnon diffusion length exhibits the depth inside YIG over which the thermal gradient is crucial for the SSE.

Our model indicates that the temporal evolution of the SSE depends on the thermal gradient in the YIG. It is important to notice that the magnon diffusion times are neglected in our model  because their group velocities are much higher than the group velocities of phonons. Further, thermal magnons up to a depth of a few hundreds of nanometer in YIG are effectively contributing to the SSE. The typical effective magnon diffusion length of 700\,nm agrees with the recent theoretical calculations \cite{Kovalev2012,Hoffman2013}. These findings rule out the possibilities of the parasitic interface effects involved in the SSE.

In conclusion, we have presented time-resolved measurements of the spin Seebeck effect in YIG$\vert$Pt bilayers performed by  the laser heating experiments. Our findings reveal that the rise time of the SSE is sub-microsecond fast, and the SSE signal attains its maximum within a few microseconds though the temperature in the system establISHEs in milliseconds. The time scale of the SSE is independent of the strength of the heating source. From our model of the magnon diffusion in thermal gradients, we find that the SSE is governed by the diffusion of the thermal magnons from the interface toward the bulk. Moreover, the establishment of the thermal gradient in the YIG film close to the interface determines the time-scales of the SSE. Our model estimates a typical diffusion length for thermal magnons to be around 700\,nm in the YIG$\vert$Pt system. Our results provide a very important piece of information about the time scales of the spin Seebeck effect that shed light on the underlying physics which might contribute to the development of future applications of spin-caloritronics.

The authors thank F. Heussner for the computational support, V. Lauer for sample-fabrication support, and T. Br\"acher, T. Meyer, and G. A. Melkov for valuable discussions. We acknowledge financial support by Deutsche Forschungsgemeinschaft (SE 1771/4-1) within Priority Program 1538 ``Spin Caloric Transport", and the technical support from the Nano Structuring Center, TU Kaiserslautern.

\end{document}